\documentstyle[prl,twocolumn,aps,graphicx]{revtex}
\tighten 

\begin{document}
\draft

\wideabs{
\title{T-shaped spin filter with a ring resonator}
\author{A. A. Kiselev\cite{email} and K. W. Kim}
\address{Department of Electrical and Computer Engineering,
North Carolina State University, Raleigh, NC 27695-7911}

\maketitle
\begin{abstract}
A planar ballistic structure is predicted to be highly effective
in filtering electron spin from an unpolarized source into two
output fluxes with the opposite and practically pure spin
polarizations. The operability of the proposed device relies on
the peculiar spin-dependent transmission properties of the
T-shaped connector in the presence of the Rashba spin-orbit
interaction as well as the difference in the dynamic phase gains
of the two alternative paths around the ring resonator through
upper and lower branches for even and odd eigenmodes.
\end{abstract}

}


The high-vacuum experiments of early 1920s performed by Stern and
Gerlach demonstrated that the trajectory of neutral silver atom is
affected by the spin eigenvalue of its unpaired electron in the
inhomogeneous magnetic field \cite{stern_gerlach}. The observed
effect is typically weak. With all recent attention to the spin
related phenomena \cite{spintronics}, the solid-state
implementation of similar principles is of considerable interest.
Propelled by the modern nanotechnology, the practically unlimited
ability to modify the matter properties suggests a conceptual
possibility to design a miniature spin filter of superior quality
and effectiveness.

In this Letter we propose a planar three-terminal ballistic
solid-state device that can filter an unpolarized incident
electrons into two output channels according to the their spin
orientation. The Stern-Gerlach nonuniform magnetic field is
substituted by the intrinsic spin-orbit (SO) interaction,
naturally present in the low-symmetry setups (for example, in
semiconductor heterostructures) and acting on the electron spin
as an effective momentum-dependent magnetic field. Various
suggestions, how to utilize the SO effect, have recently been
published in a number of spin-related proposals
\cite{datta,aronov,bulgakov,governale}. This approach is a
potentially big advantage over the external sources of the strong
magnetic field that are bulky and/or require extensive additional
circuitry (though incorporation of the ferromagnetic films into
the layered structure has also been considered \cite{cacho}). We
have analyzed several geometries, mainly symmetric and asymmetric
2D ballistic rings. Important property of this type of structures
is the possibility of at least two alternative paths for the
incoming electron that can lead to the interference and
noticeable manifestations of the intrinsic SO effect in the
spin-dependence of the transmission probabilities. The special
geometry provides high effectiveness with up to $70\%$ of the
incident unpolarized flux filtered into two output fluxes with
practically pure spin states.


The structure under consideration consists of an area of complex
geometry formed from the two-dimensional electron gas using,
e.g., an electrostatic split-gate technique, and connected to the
{\it exterior} by a number of quasi-one dimensional (1D) wires.
The simplest model 2D ($x$--$y$ plane) spin-independent electron
Hamiltonian is complimented by the spin-related term (symmetrized
to keep it Hermitian that is denoted here by curly brackets
$\{\ldots\}$):
\begin{equation}\label{hamiltonian}
H=\hat{1}\frac{\hbar^2}{2m}(k_x^2+k_y^2)+\left\{\eta(x,y)(\hat{\sigma}_xk_y
-\hat{\sigma_y}k_x)\right\}.
\end{equation}
Equation~(\ref{hamiltonian}) defines the problem along with the
boundary conditions on the two-component spinor wavefunction
$\hat{u}=0$ at the structure edges (hard walls). The third
dimension (i.e., the $z$ coordinate) is strongly quantized. The
SO interaction can be caused, in particular, by the asymmetry in
the $z$-confining potential (the so-called Rashba \cite{rashba} or
structure-asymmetry-induced term) with its strength given by the
coefficient $\eta$, $\hat{\sigma}_\alpha$ are the Pauli matrices
($\alpha=x,y,z$), and $\hat{1}$ is a $2\times2$ unity matrix. The
$\eta$ in the first approximation is proportional to the
$z$-component of the local electric field in the structure layer
and depends on the properties of the semiconductor which forms 2D
layer. For the practical means, this asymmetry in the confining
potential can be formed and/or manipulated {\it locally} by the
special control electrode(s) located over/under the structure
\cite{grundler}. For simplicity, we ignore other potential sources
of SO interaction in our system (that nevertheless exist and
compete in real systems) \cite{lommer88,krebs}. We also leave
aside any other phenomena beyond explicitly expressed in
Eq.~(\ref{hamiltonian}) (especially many-body interactions and
all types of relaxation processes). We concentrate here
exclusively on the stationary setup.


Transmission and reflection properties of the arbitrarily complex
linear system connected to the exterior via 1D wires can be
condensed into a finite-size scattering matrix ${\bf S}$
\cite{scattering_matrix}, consisting of the reflection
coefficients $r_{ii}$ (in the $i$-th channel) and transmission
coefficients $t_{ij}$ (describing propagation of the particle
from the $j$-th into the $i$-th channel). We choose to enumerate
as distinguishable channels (or terminals) all energetically
allowed electron fluxes through different 1D subbands even of the
same wire. Thus, the number of the channels can differ from the
number of the attached leads. Having said that, we will be
primarily interested in the situation when all connecting wires
are identical and the incident electron energy permits
transmission solely through the ground subband. For example, in
the hard-wall approximation the ground subband wavefunction in
the ideal lead of width $w$ is a plane wave
$\Psi^\pm=\psi_0(y)\exp(\pm ikx)$ with an envelope
$\psi_0(y)=(2/w)^{1/2}\cos(\pi y/w)$ that is symmetric in the
perpendicular direction and corresponds to the quantization
energy $E_0=\hbar^2\pi^2/2mw^2$ (these formulae are given in
coordinates describing the channel along the axis $x$). The total
energy of the state in the lead is $E=E_0+E_{\rm kin}$ where
$E_{\rm kin}=\hbar^2k^2/2m$. For the kinetic energy interval
$0<E_{\rm kin}<3E_0$ the propagation in the arms is possible only
through the ground subbands.

In the case of a spinless particle, the $r_{ii}$, $t_{ij}$ are
scalars. By taking electron spin into consideration, the number of
transmission channels effectively doubles. If the spin-dependent
interactions are not overly strong, and especially when they are
spatially localized to the {\it interior} of the system (that is
near the control electrodes in our case), it is convenient to just
``upgrade'' transmission and reflection coefficients of the ${\bf
S}$ into $2\times2$ submatrices, thus keeping pairs of the
channels differing only by the spin orientation together. Each
$2\times2$ coefficient in this case can be conveniently expanded
as
\begin{equation}\label{expansion}
\hat{x}=\hat{1}x_1+i\sum_\alpha \hat{\sigma}_\alpha x_\alpha
\mbox{~~($x=r_{ii}$ or $t_{ij}$).}
\end{equation}


With the total flux ${\bf F}\propto {\bf U}^+{\bf U}$, the
requirement of the flux conservation for ${\bf U}_{\rm out}={\bf
S}{\bf U}_{\rm in}$, corresponding to an arbitrary column ${\bf
U}_{\rm in}$ (symbolically representing coherent incident waves
$\hat{u}_i$ coming through all channels), can only be secured if
\begin{equation}\label{flux_conservation}
{\bf S}^+{\bf S}={\bf 1}.
\end{equation}

We perform now the symmetry analysis of this system. For a broad
class of problems including the one defined by
Eq.~(\ref{hamiltonian}), time-reversal invariance (with the
operator $\hat{T}=-i\hat{\sigma}_yK$ where $K$ is the complex
conjugation) establishes the following relation on the scattering
matrix ${\bf S}$
\begin{equation}\label{time_inversion}
\hat{\sigma}_y{\bf S}^*\hat{\sigma}_y{\bf S}={\bf 1}.
\end{equation}
This equation should be regarded as a symbolic one, with the Pauli
matrix multiplications applied to each submatrix $\hat{r}_{ii}$,
$\hat{t}_{ij}$ separately.

Combined with Eq.~(\ref{flux_conservation}), this relation can be
converted into a more practical form $\hat{\sigma}_y{\bf
S}^*\hat{\sigma}_y={\bf S}^+$, that immediately results into
\begin{equation}\label{r&t_components}
r_{ii,\alpha}=0, \ t_{ij,1}= t_{ji,1},\
t_{ij,\alpha}=-t_{ji,\alpha},
\end{equation}

Additional structure symmetry elements can provide further
relations on the components of ${\bf S}$. For example, for the
system symmetric in respect to the reflection $y\leftrightarrow
-y$ ($S_y$), the operator $-i\hat{\sigma}_y S_y$ commutes with the
model Hamiltonian giving
\begin{equation}\label{y_reflection}
\hat{\sigma}_yS_y({\bf S})\hat{\sigma}_y={\bf S}.
\end{equation}
The effect of reflection $S_y$ on the scattering matrix is set by
the permutation of the pairs of indices
$(i,j,\ldots)\leftrightarrow(k,l,\ldots)$ corresponding to the
symmetrically located channels. In components, that leads to
relations
\begin{eqnarray}\label{y_reflection_r&t_components}
r_{ii,1}= r_{kk,1},&& \ t_{ij,1}= t_{kl,1}, \
t_{ij,x}= -t_{kl,x},\nonumber\\
&& \ t_{ij,y}= t_{kl,y}, \ t_{ij,z}= -t_{kl,z}.
\end{eqnarray}
As a consequence, together with the relations of
Eq.~(\ref{r&t_components}) this also suggests $t_{ik,y}=0$ for
terminals $i\leftrightarrow k$. Another important case comes up
when channels $i$ and $j$ are positioned along the reflection
plane and as a result they reflect to themselves. For this setup
$t_{ij,x}= t_{ij,z}=0$.

Similar relations take place for the reflection $x\leftrightarrow
-x$ ($S_x$).
As for the $z\leftrightarrow -z$ ($S_z$), it changes the sign of
the intrinsic electric field responsible for the Rashba SO
interaction, thus establishing relations between scattering
matrices ${\bf S}(\eta)$ and ${\bf S}(-\eta)$ as
\begin{equation}\label{z_reflection}
\hat{\sigma}_z{\bf S}(-\eta)\hat{\sigma}_z={\bf S}(\eta).
\end{equation}
These simple qualitative considerations suffice for the purposes
of the present Letter \cite{Ya_i_Yulik}.


An elementary channel flux $F\propto\hat{u}^+\hat{u}$ with the
$100\%$ polarization is conveniently given by the spinor column
$\hat{u}=(u_\uparrow,u_\downarrow)$.
The relative magnitude and phase between $u_\uparrow$ and
$u_\downarrow$ characterize orientation of the spin; vector
$\bbox{P}$ defined by three components $P_\alpha$ in the $xyz$
coordinate system is called the polarization vector, $P_\alpha
F\propto\hat{u}^+\hat{\sigma}_\alpha\hat{u}$. For the arbitrary
spinor $\hat{u}$ the absolute value $|\bbox{P}|=1$. Being pumped
into the channel $j$, the $\hat{t}_{ij}\hat{u}$ part of the
incident flux will seep through the structure into the $i$ output
channel. In the general case of spin-dependent interactions
present in the system, $\hat{t}_{ij,\alpha}\neq0$ and the spin
will rotate from its original orientation. Moreover, the {\it
magnitude} of the transmitted flux will depend on the spin
orientation of the incident flux.

Partially polarized electron fluxes can be mimicked  by a number
of independent (not phase coherent) elementary fluxes $\hat{u}_q$
through the same channel that are just additive in the case of our
linear system
\begin{equation}\label{sum_of_fluxes}
F=\sum_q F_q,\ \bbox{P} F=\sum_q \bbox{P}_q F_q.
\end{equation}
To present the unpolarized input flux in channel $j$, we use, for
example, two elementary fluxes $\hat{u}_1=(1,0)$ and
$\hat{u}_2=(0,1)$. Now it is very easy to evaluate $F,\bbox{P}$
for the electron flux, transmitted into channel $i$. Indeed,
\begin{equation}\label{total}
F\propto|t_{ij,1}|^2+|t_{ij,x}|^2+|t_{ij,y}|^2+|t_{ij,z}|^2,
\end{equation}
with the $x$-component of spin polarization
\begin{equation}\label{polarization}
P_xF\propto2{\rm Im}(t_{ij,1}t_{ij,x}^*+t_{ij,y}^*t_{ij,z})
\end{equation}
and $P_y$, $P_z$ obtained by the cyclic permutation of indices.
This equation shows, in particular, that if only one component of
$\hat{t}_{ij}$ is present, polarization of the transmitted
electron flux is zero for unpolarized input.

With the polarization vector $\bbox{P}$ of the transfered flux
expressed that way, we can come to another very important
consequence with the help of Eqs.~(\ref{flux_conservation}) and
(\ref{time_inversion}): the system with the Hamiltonian of
Eq.~(\ref{hamiltonian}) and just two connecting terminals cannot
polarize transmitted electron flux. This conclusion is generally
relaxed for the structures with three or more terminals (see also
\cite{two_terminals} in relation to this matter). We restrict
ourselves here to three-terminal devices.

\begin{figure}
\begin{center}
\includegraphics*[bb=94 192 616 528,width=45mm]{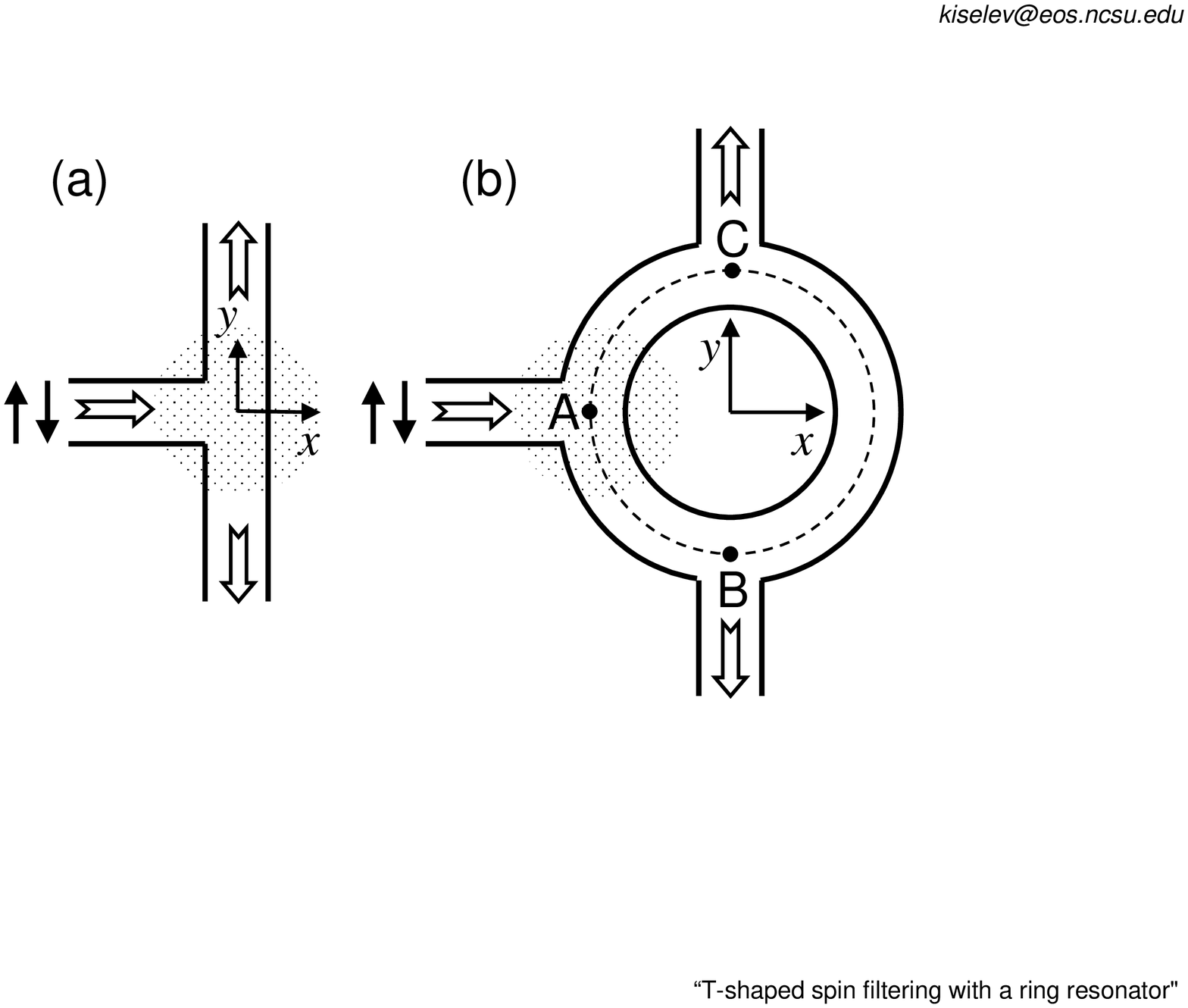}
\end{center}
\caption{\label{fring_sketch} (a) T-shaped spin filter with SO
interaction induced by the control electrode(s) placed at the
intersection; (b)~the same structure with an attached ring
resonator.}
\end{figure}

Quickly, we recollect now the results \cite{Kiselev_APL2001},
obtained for the simple T-shaped structure, formed by the
confining potential of two intersecting 1D channels [see
Fig.~\ref{fring_sketch}(a)]. The SO interaction is formed in the
2D electron gas layer by gate electrodes and localized to the
intersection area only. As the unpolarized electrons enter the
input arm ($-x$), the quantities of interest are the magnitudes
and polarizations of the two output fluxes through the $\pm y$
arms. With the $S_y$-symmetric T-shaped structure providing
symmetry relations of Eqs.~(\ref{y_reflection}) and
(\ref{y_reflection_r&t_components}), the total transmitted fluxes
of the two output channels are the same but their polarizations
along the $x$ and $z$ axes are opposite ($P_y$ is the same). The
transmission varies vary gradually with the incident energy but
has low dips in the vicinity of the resonances with the
quasi-localized 0D electron levels at the intersection
(corresponding to the total reflection in the case of no SO). The
SO interaction forms also a fine structure near these resonances.
In the resonance regions where the difference between the incident
energy and the energy of the quasi-localized 0D state becomes
comparable to the SO term, high values of polarization take place.
Actually, the smaller the SO term is, the narrower these regions
of high polarization and the higher (closer to $100\%$) the
polarization values become. Note, though, that this happens at
the expense of diminishing total transmission.

\begin{figure}
\begin{center}
\includegraphics*[bb=58 98 546 762,width=57mm]{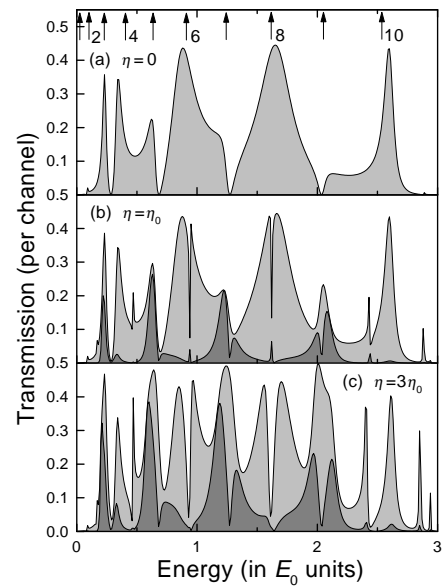}
\end{center}
\caption{\label{fring_data} Transmission of incident electron flux
into output channel of the symmetric T-shape structure with a ring
resonator as a function of electron kinetic energy. (a) No SO
interaction at the input T-shaped connector, $\eta=0$; (b)
$\eta=\eta_0$; (c) $\eta=3\eta_0$ (stronger SO interaction). The
light area represents total transmitted flux and the dark area
shows its polarized part. The numbered arrows at the top denote
approximate eigenmode energies $E_{R,l}$ in the ring.}
\end{figure}

The same T-shaped spin filter with an attached ring resonator is
shown in Fig.~\ref{fring_sketch}(b). In Figure~\ref{fring_data},
we present the energy dependence of the total transmitted flux $F$
(light area) and its polarized part $|\bbox{P}|F$ (dark) into one
of the two symmetric output channels (the flux conservation
ensures that $F\leq 0.5$). The data shown are obtained for
$\eta=0$ (no SO), $\eta_0$ (relatively small SO), and $3\eta_0$
(stronger SO) that are given in graphs (a)--(c), respectively.
Coefficient $\eta_0=6$~$\mu$eV$\cdot\mu$m is a reasonable basic
value for the Rashba constant in InAs/InGaAs heterostructures
\cite{lyanda}, electron effective mass $m=0.023 m_0$ where $m_0$
is the free electron mass, the channel width $w=0.1$~$\mu$m, and
the ring radius $R=0.2$~$\mu$m [dashed circle in
Fig.~\ref{fring_sketch}(b)]. This set of parameters gives the
ground subband quantization energy $E_0=1.6$~meV; thus, $3\eta_0$
is equal to $0.11\times E_0w$. We (arbitrarily) assume the SO
coefficient $\eta(r)$ to have the spatial dependence
$\eta(r)=\eta/[1+e^{(r-r_0)/\Delta r}]$, where $\eta$ and
$r_0=0.1$~$\mu$m are defined by the back and front electrode
potentials and their sizes, $\Delta r=0.025$~$\mu$m is included
to account qualitatively for the fringe fields,
$r=[(x+R)^2+y^2]^{1/2}$. For a numerical solution, we have
applied the recursive method of Usuki {\it et al.} \cite{usuki}.

In the absence of the SO interaction [Fig.~\ref{fring_data}(a)],
only $t_{ij,1}$ terms are present in the transmission
coefficients and the polarization of the output fluxes repeats
that of the incident fluxes; thus, it is zero for the unpolarized
input flux. Energy dependence of the transmission probability
consists of a number of resonant peaks, corresponding to the
eigenstates in the ring. Strictly speaking, eigenstates in the
ideal ring should be defined via the Bessel functions. Under the
assumption $w\ll 2\pi R$, energies of several first states can be
approximated by $E_{R,l}=k_l^2/2m$ (counted from the quantization
energy $E_0$ in the ideal channel of width $w$), where $k_l=l/R$
and $l$ is an integer. Thus, except for the ground state $l=0$,
the eigenstates in the ring are quadruply degenerate
--- twice because of the orbital motion $\pm l$, and twice due to
the spin. Presence of the attached terminal leads inflicts a
finite lifetime of the ring eigenmodes (and leads to some
renormalization of the energies).

The arrows in the upper part of the graph indicate several first
energies $E_{r,l}$. One can immediately note that for {\it even}
orbital numbers $l$, transmission peaks are well defined, while
for {\it odd} $l$, the peaks are suppressed and the main feature
here is a dip in the transmission probability. This can
qualitatively be explained as follows. An electron wave
transferring to, say, upper output channel passes $1/4$ of the
ring if moving through the upper branch $A\to C$ and $3/4$ for
the lower branch $A\to B\to C$. Thus, the difference in the
gained dynamic phase \cite{topological} is $\pi l$ that gives a
multiplier of $+1$ for even and $-1$ for odd modes that secures
either constructive or destructive interference at the output
connector. Deviation from the described scenario, especially for
a couple of the first eigenmodes can be explained by the leakage
of the electron wave in the lower branch into the $-y$ lead (the
leakage is stronger for smaller $E_{\rm kin}$), thus, severely
disturbing the balance of the wave amplitudes arriving via two
paths.

Now it is easy to understand the structure of the transmission
spectra in the presence of the SO interaction
[Figs.~\ref{fring_data}(b) and \ref{fring_data}(c)]. All peaks
are split into two subpeaks because the four-fold degeneracy of
the eigenstates in the ring is lifted into two pairs by the SO
mixing \cite{two_pairs}. Taking into consideration the symmetry
relations of Eqs.~(\ref{y_reflection}) and
(\ref{y_reflection_r&t_components}) for the $S_y$-symmetric
T-shaped connector, one can conclude that the explanation given
above is still valid for the $t_{ij,1}$, $t_{ij,y}$ components of
the transferred flux, but for the $t_{ij,x}$ and $t_{ij,z}$,
where the sign for the waves transmitted into the upper and lower
branches are opposite, constructive interference takes place for
{\it odd} eigenmodes in the ring. That is indeed very important
conclusion since constructive interference of the $x-z$ polarized
and unpolarized electron fluxes take place for {\it drastically
different} incident energies. This explains really high
polarization for the energies corresponding to the odd modes ---
the polarization reaches $100\%$ where the dark-area
($|\bbox{P}|F$) and the light-area ($F$) contour boarders adjoin.

Another important observation concerns the $P_y$-component of the
transmitted flux. It is the same for both $+y$ and $-y$ output
fluxes in the symmetric structure. That means that for the odd
eigenmodes the corresponding flux component interferes
destructively, just like the unpolarized flux fracture. Thus,
$P_y\approx0$ and polarizations of two output fluxes are not only
extremely high, but they are also {\it practically exactly}
opposite for two output arms, which is indeed an optimal
situation for the filter.

So far we have implicitly considered monoenergetic incident
electrons. For the described effect to be observable in a more
realistic setup, the energy distribution of the incident electrons
should not substantially exceed the energy gap between adjacent
ring eigenmodes and be aligned in such a way that to overlap
primarily with only one subpeak of the SO split transmission
peak, as the spin polarization is approximately opposite for two
subpeaks. These requirements could be fulfilled by adjusting the
Fermi levels of contact reservoirs and performing experiments in a
proper temperature range.




In summary, we have proposed a device consisting of the T-shaped
spin filter with an attached ring resonator. Matching incident
electron energies to some of the eigenmodes in the ring, we have
been able to achieve both a superior spin selectivity and high
transmission efficiency.


This work was supported by the ONR and
DARPA.



\begin{references}
\bibitem{topological} Due to the localization of the SO interaction
in a small area at the left connector, topological phase of
\cite{aronov} is not involved here.
\bibitem{two_pairs}
For the particular case depicted in Fig.~\ref{fring_sketch}(b)
with the Rashba SO interaction localized at the left connector,
these eigenstate pairs are: $(l,1/2)$ (the orbital number $l$ and
the spin oriented along axis $x$  with $s_x=1/2$) and $(-l,-1/2)$;
the second pair is $(l,-1/2)$ and $(-l,1/2)$.
\end{references}
\end{document}